\documentclass[11pt]{article}
\usepackage[dvips]{color}
\usepackage{epsfig}
\usepackage{amsmath}
\usepackage{graphicx}
\date{}
\newcommand{\christoffelg}[3]{\{ g \}^{{#1}}_{{#2}\hspace*{0.2em}{#3}}}
\newcommand{\christoffelX}[3]{\{ X \}^{{#1}}_{{#2}\hspace*{0.2em}{#3}}}
\newcommand{\christoffelf}[3]{\{ f \}^{{#1}}_{{#2}\hspace*{0.2em}{#3}}}
\begin{document}
\title{{\bf $SO(4,2)$ and derivatively coupled dRGT massive gravity}}


\author{Nafiseh Rahmanpour$^1$\thanks{%
e-mail: n\_rahmanpour@sbu.ac.ir},\,\, Nima Khosravi$^1$\thanks{%
e-mail: n-khosravi@sbu.ac.ir}\,\, and Babak Vakili$^2$\thanks{%
e-mail: b.vakili@iauctb.ac.ir }
\\\\
$^1${\small {\it Department of Physics, Shahid Beheshti University, G.C., Evin, Tehran 19839, Iran}}\\
$^2${\small {\it Department of Physics, Central Tehran Branch,
Islamic Azad University, Tehran, Iran}}}

\maketitle

\begin{abstract}
In this paper we study the possibility of assigning a geometric
structure to the Lie groups. It is shown the Poincar\'{e} and Weyl
groups have geometrical structure of the Riemann-Cartan and Weyl
space-time respectively. The geometric approach to these groups can
be carried out by considering the most general (non)metricity
conditions, or equivalently, tetrad postulates which we show that
can be written in terms of the group's gauge fields. By focusing on
the conformal group we apply this procedure to show that a
nontrivial 3-metrics geometry may be extracted from the group's
Maurer-Cartan structure equations. We systematically obtain the
general characteristics of this geometry, i.e. its most general
nonmetricity conditions, tetrad postulates and its connections. We
then deal with the gravitational theory associated to the conformal
group's geometry. By proposing an Einstein-Hilbert type action, we
conclude that the resulting gravity theory has the form of
quintessence where the scalar field derivatively coupled to massive
gravity building blocks.

\vspace{5mm}\noindent\vspace{0.8mm}
\end{abstract}

\section{Introduction}
It is well known that present-day understanding of structure of
space-time is based on the Einstein's theory of general relativity
(GR) which some principles such as principle of equivalence, general
covariance and isometry group, form its symmetry properties. In its
first formulation done by Einstein gravity is linked to the
geometrical properties of space-time which in turn, is fully
determined by the (dynamical) metric tensor $g_{\mu \nu}$. In
addition, a metric compatible torsion-free connection, usually
called Levi-Civita connection will appear which may be extracted
from metric by the metricity condition $\nabla_\rho g_{\mu\nu}=0$.
Then, equations of motion for metric are the result of variation of
Einstein-Hilbert action with respect to the metric. However, in
general, these two objects are completely independent from each
other and a manifold may be equipped with metric and connection as
independent variables. Although the variation of the
Einstein-Hilbert action of GR (in Palatini formalism) with respect
to connection gives no more information other than the metricity
condition mentioned above, this is not the case for the generalized
theories of gravity that their actions are more complicated than GR
\cite{Sot}.

In addition, the formulation of gravity seems to be somehow
different from other known interactions which are described by
Yang-Mills gauge theories. The action of all interactions but
gravity is constructed by means of the invariance under the act of
some internal Lie groups, some example are $U(1)$, $SU(2) \times
U(1)$ and $SU(3)$ gauge theories for quantum electrodynamics,
electroweak and quantum chromodynamics respectively. These gauge
theories have a natural formulation within the fiber bundle
framework. GR may also be formulated within this framework as a
gauge theory, however, the most important difference between
gravitation and other gauge theories is due to the solder form which
relates the bundle and base manifold.

The term of gauge first appeared in the works of Hermann Weyl
\cite{Weyl}, in which he attempted to unify gravitation and
electromagnetism, see also \cite{Ful}. Weyl formulated his theory in
a geometry without metricity condition. Therefore, the first
prototype model of gauge theories was indeed a gravitational theory,
although after that the other forces have been formulated and known
as a gauge theory. Another pioneering attempt to formulate gravity
in the framework of a gauge theory has been done by \'{E}lie Cartan
who tried to generalize the space-time structure by assuming the
metric and the non-symmetric affine connection as independent
quantities \cite{Cartan}. Since then, many efforts have been made in
this regard and their results have been followed and developed by a
number of works, for instance \cite{Hehl}-\cite{Kop}. A notable
point about the gravity as a gauge theory is that instead of the
Yang-Mills theories of internal symmetries, here one should deal
with some external (space-time) gauge symmetry such as Poincar\'{e},
Conformal and de-Sitter gauge theories; in which the usual
diffeomorphisms and gauge transformations are linked to each other
in the corresponding fiber bundle.

In gauge theories, we can distinguish two approaches. On one hand,
as mentioned above, a gauge theory is considered as a theory in
which the action is invariant under a continuous symmetry Lie group
that depends on space-time. Then, gauge symmetries introduce gauge
fields to the theory which mediate a force. The gauge fields are
massless unless spontaneous symmetry breaking occurs. In this
approach, one first chooses a gauge group and then tries to
construct a Lagrangian in such a way that it respects to the gauge
symmetry. On the other hand, we can begin with a Lagrangian with a
symmetry and then try to localize it. As an example, in the local
Poincar\'{e} gauge theory, the gauge fields are obtained by
requiring invariance of the Lagrangian density under the local
Poincar\'{e} transformations. From the point of this view, it can be
shown that there is a geometric interpretation corresponding to the
present symmetry Lie group. In the cases of the Poincar\'{e} and
Poincar\'{e}+Scale gauge groups, for instants, the corresponding
geometries are Riemann-Cartan and Weyl-Cartan respectively
\cite{Ali}.

In this regard, we may still go further and raise a question: Can a
generalized theory of gravity be developed beyond the geometric
structure derived from the gauge theory? Inspired by some recent
works, \cite{Hassan} for instance, in which a bimetric theory of
gravity is derived from the $SO(1,5)$ gauge theory and \cite{Tomi}
in which the relation between extended Gauss-Bonnet gravities and
Weyl geometry has been investigated, it seems that this question has
a positive answer, see also \cite{tomi1} for a review on the
geometrical foundation of gravity. In this paper, we are going to
deal with this question in the framework of the $SO(4,2)$ conformal
group. This group has already been used to study of some infrared
modifications of gravity \cite{Seahra}. To do this, we begin with a
Lie group $G$ for which a distinguished differential 1-form
$\omega$, usually called Maurer-Cartan form exists that carries the
basic infinitesimal information of the structure of the group $G$.
Maurer-Cartan form is actually a Cartan connection of the Cartan
geometry and is a special case of the principal connection. This
means that in the gauge gravity models, the Maurer-Cartan form may
be identified with the vector potential $A$. After writing the
Maurer-Cartan structure equations corresponding to the selected
group we use the building blocks and symmetries of the groups to
propose a geometric interpretation. By examining this strategy on
the Poincar\'{e} and Weyl groups we get the corresponding geometric
structure and then we apply the mentioned method on the conformal
group. Interestingly, we encountered a three-metric geometry
constructed from the group's gauge fields with non-trivial metricity
conditions and showed that the gravity theory corresponds to it has
an additional scalar field which is coupled to metric via massive
gravity building blocks.

The paper is organized as follows. In section 2, we will provide the
general framework of our work and describe how the Cartan connection
can be written in terms of the gauge fields and structure constants
of a Lie group. The general form of the Maurer-Cartan structure
equations is also introduced in this section and for two special
cases of Poincar\'{e} and Weyl groups the detail are presented.
Section 3 is devoted to our main goal, that is, the conformal group
and its geometrical properties which are deduced based on the method
developed in section 2. In section 4, we presented a gravity theory
according to the space-time geometry of the conformal group.
Finally, we summarize the results in section 5.

\section{Lie groups and their associated geometry}
In this section we consider a Lie group $G$ with $\textbf{g}$ being
its Lie algebra. The generators {$T_a$} of the group form the closed
commutation relations $[T_a,T_b]=f^c_{ab}T_c$, where
$f^c_{ab}=-f^c_{ba}$ are the structure functions. With each
generator we associate a dual 1-form $\theta^a$ in such a way that
$<\theta^a,T_b>=\delta^a_b$. While the elements of the set $\{T_a\}$
are linearly independent left-invariant vector fields at each point
in $G$, the elements of the set $\{\theta^a\}$ form the basis for
the left-invariant 1-forms. The Maurer-Cartan form
$\omega=\theta^aT_a$ is a $\textbf{g}$-valued 1-form which satisfies
the Maurer-Cartan structure equation

\begin{equation}\label{A}
d\omega+\omega \wedge \omega=0,
\end{equation}which can be written in the following equivalent form

\begin{equation}\label{B}
d \theta ^c+\frac{1}{2} f_{ab}^{c} \theta^{a} \wedge \theta^{b}=0.
\end{equation}We may generalize the connection in the Maurer-Cartan equation
by adding the curvature to it, as

\begin{equation}\label{C}
d\omega+\omega \wedge \omega=\Omega,
\end{equation}where $\Omega$ is the curvature 2-form. In the
following two subsections by considering the Poincar\'{e} and Weyl
groups, we will review how starting from the Maurer-Cartan equations
leads us to the corresponding geometric structure of each of these
groups.

\subsection{Poincar\'{e} group}
The Poincar\'{e} group consists of two Lie groups: the group of
Lorentz transformation with generators $M_{ab}$ and the group of
translation with generators $P_a$. It can be verified that they
satisfy the following Lie algebra

\begin{eqnarray}\label{D}
\left\{
\begin{array}{ll}
\left[M_{ab}, M_{cd}\right]=\eta_{ad} M_{bc}+\eta_{bc} M_{ad}+\eta_{ac} M_{db}+ \eta_{bd} M_{ca},\\\\
\left[P_{a}, M_{bc}\right]=\eta_{ba} P_{c}- \eta_{ca} P_{b},\\\\
\left[P_a,P_b\right]=0.
\end{array}
\right.
\end{eqnarray}Now, the Maurer-Cartan 1-form takes the form $\omega=\theta^a P_a+
\omega^{a}_{\,\,\,b} M^{b}_{\,\,\,a}$, where
$\theta^a=\theta^{\,\,\,a}_{\mu}dx^{\mu}$ and
$\omega^a_{\,\,\,b}=\omega^{\,\,\,a}_{\mu \,\,\, b}dx^{\mu}$ are the
non-coordinate (tetrad) basis and spin connection 1-forms
respectively \cite{Yepez}. The Maurer-Cartan equations then read

\begin{eqnarray}\label{E}
\left\{
\begin{array}{ll}

T^a=d \theta^a + \omega^{a}_{\,\,\,c} \wedge \theta^{c}, \\\\
\Omega^{a}_{\,\,\,b}=d\omega^{a}_{\,\,\,b} + \omega^{a}_{\,\,\,c}
\wedge \omega^{c}_{\,\,\,b},
\end{array}
\right.
\end{eqnarray}in which, in terms of the Cartan's terminology, $T^a$
is the torsion 1-form and $\Omega^a_{\,\,\,b}$ denotes the (Riemann)
curvature 2-form. By means of the tetrad field one may now construct
a metric $g_{\mu \nu}$ as

\begin{equation}\label{F}
g_{\mu\nu}=\theta^{\,\,\,a}_{\mu} \theta^{\,\,\,b}_{\nu}\eta_{ab}.
\end{equation}In addition, considering $\omega^a_{\,\,\,b}$ as a spin connection,
namely as the non-coordinate based connection corresponding to the
Levi-Civita connection ${\Gamma}^{\nu}_{\rho \mu}$, leads to

\begin{eqnarray}\label{G}
0&=&\nabla_\rho g_{\mu\nu} \nonumber \\
&=&D_{\rho}(\eta_{ab}\theta^{\,\,\,a}_{\mu}\theta^{\,\,\,b}_{\nu})\nonumber\\
&=&(\eta_{ab}\theta^{\,\,\,b}_{\nu})D_{\rho}\theta^{\,\,\,a}_{\mu}+
(\eta_{ab}\theta^{\,\,\,a}_{\mu})D_{\rho}\theta^{\,\,\,b}_{\nu}
\end{eqnarray}in which by $D_{\rho}$ we mean the covariant
derivative operating on both internal and spacetime indices. One
possible solution to the above relation is $D_\rho
\theta^{\,\,\,a}_{\mu}=0$ which is

\begin{equation}\label{H}
D_\rho \theta^{\,\,\,a}_{\mu}=\partial_\rho \theta^{\,\,\,a}_{\mu}+
\omega^{\,\,\,a}_{\rho \,\,\, b}
\theta^{\,\,\,b}_{\mu}-{\Gamma}^{\nu}_{\rho \mu}
\theta^{\,\,\,a}_{\nu}=0.
\end{equation}This relation in the literature is mentioned as the tetrad postulate and is
indeed the result of consistency of frames \cite{Yepez}. Although,
as it is shown here, it can be derived from metricity condition. It
can easily be shown that this argument can be reversed, that is,
starting from the tetrad postulate we may get the metricity
condition:

\begin{equation}\label{I}
\nabla_\rho g_{\mu\nu}=0 \Leftrightarrow D_\rho
\theta^{\,\,\,a}_{\mu}=0.
\end{equation}
The next step is to use the equations (\ref{F}) and (\ref{H}) for
$\theta^a$ and $\omega^{a}_{\,\,\,b}$, in favor of $g$ and $\Gamma$
and substitute the results into the Maurer-Cartan equations. By
doing the calculation, while the first relation yields the torsion
tensor, the second one gives us the Riemann curvature tensor. This
demonstrates that Poincar\'{e} gauge theory has the geometrical
structure of the Riemann-Cartan space-time which in the case where
torsion vanishes reduces to the usual Riemann space-time of GR.

\subsection{Weyl group}
In addition to the transformations of the Poincar\'{e} group, the
Weyl group has an additional transformation called dilatation which
is responsible for the resize of a coordinate: $x^a\rightarrow
e^{\lambda} x^a$. It can be shown that its generator has the form:
$D=x^a \partial_a$ \cite{Juan}. The commutation relations between
the generators of the Weyl group generate the following Lie algebra
\cite{Blag}

\begin{eqnarray}\label{J}
\left\{
\begin{array}{ll}
\left[M_{ab}, M_{cd}\right]=\eta_{ad} M_{bc}+\eta_{bc} M_{ad}+\eta_{ac} M_{db}+ \eta_{bd} M_{ca},\\\\
\left[P_{a}, M_{bc}\right]=\eta_{ba} P_{c}- \eta_{ca} P_{b},\\\\
\left[P_a,D\right]=P_a,\\\\
\left[P_a,P_b\right]=\left[M_{ab},D\right]=\left[D,D\right]=0.
\end{array}
\right.
\end{eqnarray}
Following the same steps as in the previous subsection, we can
define the Maurer-Cartan 1-form as

\begin{equation}\label{K}
\omega=\theta^a P_a+ \omega^{a}_{\,\,\,b} M^{b}_{\,\,\,a}+\kappa D,
\end{equation}where $\kappa$ is a gauge field corresponding to the dilatation usually called Weyl vector.
So, the Maurer-Cartan structure equations for the Weyl group take
the form

\begin{eqnarray}\label{L}
\left\{
\begin{array}{ll}
T^a=d \theta^a + \omega^{a}_{\,\,\,c} \wedge \theta^{c}-\kappa \wedge \theta^{a}, \\\\
\Omega^{a}_{\,\,\,b}=d\omega^{a}_{\,\,\,b} + \omega^{a}_{\,\,\,c}
\wedge \omega^{c}_{\,\,\,b},\\\\
K=d\kappa,
\end{array}
\right.
\end{eqnarray}in which the dilatation curvature $K$ is introduced.
It is seen that in Weyl group one has three building blocks
$\theta$, $\omega$ and $\kappa$. While as before a metric $g_{\mu
\nu}$ may be constructed by using of $\theta$s, the field $\kappa$
leads $\omega^a_{\,\,\,b}$ not to be a usual spin connection. In
other words, in Weyl group, $\omega^a_{\,\,\,b}$ cannot be
considered as associated connection to the Levi-Civita connection.
Therefore, as a generalization for the relation (\ref{H}), we assume

\begin{equation}\label{M}
D_\rho \theta^{\,\,\,a}_{\mu}=\frac{1}{2}\kappa_\rho
\theta^{\,\,\,a}_\mu,
\end{equation}which in terms of the coordinate language reads

\begin{equation}\label{N}
\partial_\rho \theta^{\,\,\,a}_{\mu}+\omega^{\,\,\,a}_{\rho \,\,\,b}\theta^{\,\,\,b}_{\mu}-\Gamma^\nu_{\rho \mu} \theta^{\,\,\,a}_{\nu}
=\frac{1}{2}\kappa_\rho \theta^{\,\,\,a}_{\mu}.
\end{equation}
With a straightforward calculation one can show that this relation
is equivalent to the familiar non-metricity condition in Weyl
geometry \cite{Weyl1}

\begin{equation}\label{O}
\nabla_\rho g_{\mu\nu}=\kappa_\rho g_{\mu\nu},
\end{equation}which as the case of the previous subsection may be
considered as the starting point from which the tetrad postulate
(\ref{M}) can be obtained, that is

\begin{equation}\label{P}
\nabla_\rho g_{\mu\nu}=\kappa_\rho g_{\mu\nu}\Leftrightarrow D_\rho
\theta^{\,\,\,a}_{\mu}=\frac{1}{2}\kappa_\rho
\theta^{\,\,\,a}_{\mu}.
\end{equation}

If torsion vanishes the above relations result

\begin{equation}\label{Q}
\Gamma^\nu_{\rho \mu} =\christoffelg{\nu}{\rho}{ \mu} +C^\nu_{\rho
\mu},
\end{equation}where $\christoffelg{\nu}{\rho}{ \mu}$ is the
Christoffel symbol of the metric $g$ and

\begin{equation}\label{R}
C^\nu_{\rho \mu}=\frac{-1}{2}\left(\delta^\nu_\rho
\kappa_\mu+\delta^\nu_\mu \kappa_\rho-g_{\rho \mu} \kappa^\nu
\right).
\end{equation}By using Eqs. (\ref{F}) and (\ref{N}) for $\theta^a$ and
$\omega^{a}_{\,\,\,b}$, these quantities can be expressed in terms
of $g$ and $\Gamma$. The second relation of the Maurer-Cartan
structure equations then gets the curvature tensor of Weyl geometry
which means that one can consider the Weyl geometry as associated
geometry of Weyl group.

\section{Conformal group and its related geometry}

In this section we will address the issue of the relation between
the Lie groups and their corresponding geometric structure in the
framework of the conformal group. The conformal group  includes the
most general transformations preserving the ratios of the lengths.
Aside from the transformations in the Weyl group, this group has an
additional transformation which in the literature is called special
conformal transformation or co-translation with generator
$L_a=-2x_ax^b\partial_b+x^2\partial_a$ \cite{Juan}. By taking of all
possible commutation relations between the generators of the
conformal group we are led to the following Lie algebra
\cite{Juan,Blag}

\begin{eqnarray}\label{S}
\left\{
\begin{array}{ll}
\left[M_{ab}, M_{cd}\right]=\eta_{ad} M_{bc}+\eta_{bc} M_{ad}+\eta_{ac} M_{db}+ \eta_{bd} M_{ca},\\\\
\left[P_{a}, M_{bc}\right]=\eta_{ba} P_{c}- \eta_{ca} P_{b},\\\\
\left[P_a,D\right]=P_a,\\\\
\left[D, L_a\right] = L_a,\\\\
\left[L_{a}, M_{bc}\right] =\eta_{ba} L_{c}- \eta_{ca} L_{b},\\\\
\left[P_{a}, L_{b}\right] =2\left(-\eta_{ab} D+M_{ab}\right),\\\\
\left[P_a,P_b\right]=\left[M_{ab},D\right]=\left[D,D\right]=\left[L_a,L_b\right]=0.
\end{array}
\right.
\end{eqnarray}By means of the structure functions of the above Lie
algebra the Cartan connection can be written as

\begin{equation}\label{T}
\omega=\theta^a P_a+\sigma^a L_a+ \omega^{a}_{\,\,\,b}
M^{b}_{\,\,\,a}+\kappa D,
\end{equation}where the co-solder form $\sigma^a$, are the gauge fields of the
co-translations. Having all these at hand we arrive at the
Maurer-Cartan equations as

\begin{eqnarray}\label{U}
\left\{
\begin{array}{ll}
T^a=d \theta^a + \omega^{a}_{\,\,\,c} \wedge \theta^{c}-\kappa \wedge \theta^{a}, \\\\
S^a=d \sigma^a +\omega^{a}_{\,\,\,c} \wedge \sigma^{c}+\kappa \wedge
\sigma^{a},\\\\ \Omega^{a}_{\,\,\,b}=d\omega^{a}_{\,\,\,b}+
\omega^{a}_{\,\,\,c}\wedge \omega^{c}_{\,\,\,b}+\sigma^{[a} \wedge \theta_{b]},\\\\
K=d\kappa+\sigma_{a}\wedge\theta^{a},
\end{array}
\right.
\end{eqnarray}where $S^a$ may be called special conformal curvature. As we did in the previous
section, we are now ready to make a metric using the gauge fields.
However, it should be noted that with an extra field $\sigma^a$, we
will face three different possibilities to arrive a metric, that is

$\bullet$ The metric $g$: as before may be constructed using the
gauge fields $\theta^a$ as

\begin{equation}\label{V}
g_{\mu\nu}=\theta^{\,\,\,a}_{\mu} \theta^{\,\,\,b}_{\nu}\eta_{ab}.
\end{equation}

$\bullet$ The metric $f$: can be constructed using the gauge fields
$\sigma^a$ as

\begin{equation}\label{W}
f_{\mu\nu}=\sigma^{\,\,\,a}_{\mu} \sigma^{\,\,\,b}_{\nu}\eta_{ab}.
\end{equation}

$\bullet$ The metric $X$: which may be defined by a symmetric
combination of the fields $\theta$ and $\sigma$, as

\begin{equation}\label{X}
X_{\mu\nu}=\theta^{\,\,\,a}_{(\mu} \sigma^{\,\,\,b}_{\nu)}\eta_{ab}.
\end{equation}Before going any further, let us see how can we make
the metric $X$ symmetric and how its inverse may be well-defined. To
do these, we use the Deser-Van Nieuwenhuizen condition \cite{Deff}

\begin{eqnarray}\label{Y}
\theta^\mu_{\,\,\,a}\sigma_{b\mu}&=&\theta^\mu_{\,\,\,b}\sigma_{a\mu},\\
\theta^{\,\,\,a}_{\mu}
\sigma_{a\nu}&=&\theta^{\,\,\,a}_{\nu}\sigma_{a\mu},
\end{eqnarray}from which we get

\begin{eqnarray}\label{Z}
X_{\mu\nu}&=& \theta^{\,\,\,a}_{\mu} \sigma^{\,\,\,b}_{\nu}\eta_{ab}\nonumber\\
&=&\theta^{\,\,\,a}_{\nu} \sigma^{\,\,\,b}_{\mu}\eta_{ab}\nonumber\\
&=& X_{\nu\mu}.
\end{eqnarray}Also, at the first glance, definition of the inverse of the metric $X$ does not seem to
be straightforward. Namely, if one assumes

\begin{equation}\label{AA}
\left(X^{-1}\right)^{\mu\nu}=\theta_{\,\,\,a}^{(\mu}
\sigma_{\,\,\,b}^{\nu)}\eta^{ab},
\end{equation}
then
\begin{equation}\label{AB}
\left(X^{-1}\right)^{\mu\beta} X_{\mu\nu}=
\frac{1}{2}\delta^{\beta}_{\nu}+\frac{1}{4}\left(\theta^{\,\,\,a}_{\mu}\sigma_{a\nu}\theta^{\beta}_{\,\,\,c}
\sigma^{\mu
c}+\theta^{\,\,\,a}_{\nu}\sigma_{a\mu}\theta^{\mu}_{\,\,\,c}
\sigma^{\beta c}\right).
\end{equation}However, by using of the Deser-Van Nieuwenhuizen condition, this relation reduces
to $\left(X^{-1}\right)^{\mu\beta}X_{\mu\nu}=\delta^{\beta}_{\nu}$.
Also, note that the structure of the conformal group does not change
under

\begin{equation}\label{AC}
\theta  \rightarrow \sigma,\hspace{5mm} \sigma \rightarrow \theta,
\hspace{5mm} \kappa \rightarrow -\kappa,
\end{equation}or equivalently

\begin{equation}\label{AD}
g \rightarrow f,\hspace{5mm} f \rightarrow g,\hspace{5mm}
X\rightarrow X, \hspace{5mm} \kappa \rightarrow -\kappa.
\end{equation}
Let us now see if a particular geometric structure can be attributed
to the conformal group. To begin, we suggest the following tetrad
postulate-like relations

\begin{eqnarray}\label{AE}
\left\{
\begin{array}{ll}
D_\rho \theta^{\,\,\,a}_{\mu}=\kappa_\rho\left(M
\theta^{\,\,\,a}_{\mu}+N \sigma^{\,\,\,a}_{\mu}\right),\\\\
D_\rho
\sigma^{\,\,\,a}_{\mu}=\kappa_\rho\left(A\theta^{\,\,\,a}_{\mu}+B
\sigma^{\,\,\,a}_{\mu}\right),
\end{array}
\right.
\end{eqnarray}where $M$, $N$, $A$ and $B$ are some coefficients. The
requirement that these relations should satisfy the symmetries in
equation (\ref{AC}) yields $N=-A$ and $B=-M$. So, the most general
tetrad postulates which respect the mentioned symmetries are

\begin{eqnarray}\label{AF}
\left\{
\begin{array}{ll}
D_\rho \theta^{\,\,\,a}_{\mu}=\kappa_\rho\left(M
\theta^{\,\,\,a}_{\mu}-A \sigma^{\,\,\,a}_{\mu}\right),\\\\
D_\rho
\sigma^{\,\,\,a}_{\mu}=\kappa_\rho\left(A\theta^{\,\,\,a}_{\mu}-M
\sigma^{\,\,\,a}_{\mu}\right).
\end{array}
\right.
\end{eqnarray}In order to obtain the nonmetricity conditions
associated to the metrics $X$, $f$ and $g$, we do as follows. First,
it is not difficult to see that the most general nonmetricity
condition for the metric $X$ that also respects the relations
(\ref{AC}) and (\ref{AD}) is of the form

\begin{equation}\label{AG}
\nabla_\rho X_{\mu\nu}=A \kappa_\rho \left(g_{\mu\nu}- f_{\mu\nu}
\right).
\end{equation}On the other hand, by multiplying the first equation
of (\ref{AF}) by $\theta$ and the second one by $\sigma$, one gets

\begin{eqnarray}\label{AH}
\left\{
\begin{array}{ll}
\theta^{\,\,\,b}_{\nu}D_\rho
\theta^{\,\,\,a}_{\mu}=\kappa_\rho\left(M
\theta^{\,\,\,b}_{\nu}\theta^{\,\,\,a}_{\mu}-A \theta^{\,\,\,b}_{\nu}\sigma^{\,\,\,a}_{\mu}\right),\\\\
\sigma^{\,\,\,b}_{\nu}D_\rho
\sigma^{\,\,\,a}_{\mu}=\kappa_\rho\left(A\sigma^{\,\,\,b}_{\nu}\theta^{\,\,\,a}_{\mu}-M\sigma^{\,\,\,b}_{\nu}
\sigma^{\,\,\,a}_{\mu}\right),
\end{array}
\right.
\end{eqnarray}where by changing the indices as $a\leftrightarrow b$
and $\mu \leftrightarrow \nu$ read

\begin{eqnarray}\label{AI}
\left\{
\begin{array}{ll}
\theta^{\,\,\,a}_{\mu}D_\rho
\theta^{\,\,\,b}_{\nu}=\kappa_\rho\left(M
\theta^{\,\,\,a}_{\mu}\theta^{\,\,\,b}_{\nu}-A \theta^{\,\,\,a}_{\mu}\sigma^{\,\,\,b}_{\nu}\right),\\\\
\sigma^{\,\,\,a}_{\mu}D_\rho
\sigma^{\,\,\,b}_{\nu}=\kappa_\rho\left(A\sigma^{\,\,\,a}_{\mu}\theta^{\,\,\,b}_{\nu}-M\sigma^{\,\,\,a}_{\mu}
\sigma^{\,\,\,b}_{\nu}\right).
\end{array}
\right.
\end{eqnarray}By summing up the sides of the relations (\ref{AH})
and (\ref{AI}) we arrive at

\begin{equation}\label{AJ}
\nabla_\rho g_{\mu\nu}=2\kappa_\rho \left(M g_{\mu\nu}-A
X_{\mu\nu}\right),
\end{equation}
and
\begin{equation}\label{AK}
\nabla_\rho f_{\mu\nu}=2\kappa_\rho \left(A X_{\mu\nu}-M
f_{\mu\nu}\right).
\end{equation}Hence, the structure of the conformal group imposes a  3-metric geometrical structure whose nonmetricity
conditions are given in (\ref{AG}), (\ref{AJ}) and (\ref{AK}). These
are also the most general relations which respect the symmetries
(\ref{AC}) and (\ref{AD}). Now, a question may be: what is the
connection associated to these relations. To answer, let us expand,
for instance, the relation (\ref{AG}) and then apply the cyclic
permutation $\rho \rightarrow \mu \rightarrow \nu \rightarrow \rho$
on its indices. The result is

\begin{eqnarray}
\partial_\rho X_{\mu\nu}- \Gamma^\alpha_{\rho \mu}X_{\alpha\nu}-\Gamma^\alpha_{\rho \nu}X_{\alpha\mu} &=&A \kappa_\rho \left(g_{\mu\nu}-
f_{\mu\nu}\right),
\label{AL}\\
\partial_\mu X_{\nu\rho}-\Gamma^\alpha_{\mu\nu}X_{\alpha\rho}-\Gamma^\alpha_{\mu\rho}X_{\nu\alpha}&=&A \kappa_\mu \left(g_{\nu\rho}-
f_{\nu\rho}\right),
\label{AM}\\
\partial_\nu X_{\rho\mu}-\Gamma^\alpha_{\nu\rho}X_{\alpha\mu}-\Gamma^\alpha_{\nu \mu}X_{\rho\alpha}&=&A \kappa_\nu \left(g_{\mu\rho}- f_{\mu\rho}\right),\label{AN}
\end{eqnarray}
where the connection is assumed to be symmetric in its lower
indices. The combination
$\left[(\ref{AL})-(\ref{AM})+(\ref{AN})\right]$ gives

\begin{eqnarray}\label{AO}
& &(\partial_\rho X_{\mu\nu}+\partial_\nu X_{\rho\mu}-\partial_\mu
X_{\nu\rho})- (\Gamma^\alpha_{\nu \mu}-\Gamma^\alpha_{\mu\nu})
X_{\alpha\rho} - (\Gamma^\alpha_{\nu\rho}+\Gamma^\alpha_{\rho
\nu})X_{\alpha\mu}-
(\Gamma^\alpha_{\rho \mu}- \Gamma^\alpha_{\mu\rho})X_{\alpha\nu}\nonumber\\
&=&A\left[\kappa_\rho ( g_{\mu\nu}- f_{\mu\nu})+\kappa_\nu (
g_{\mu\rho}- f_{\mu\rho})-\kappa_\mu ( g_{\nu\rho}-
f_{\nu\rho})\right].
\end{eqnarray}
Multiplying both sides of the above relation by
$\frac{1}{2}\left(X^{-1}\right)^{\mu\beta}$ and assuming that the
connection $\Gamma$ is symmetric results

\begin{eqnarray}\label{AP}
\Gamma^\beta_{\nu\rho}=\christoffelX{\beta}{\nu}{\rho}-\frac{A}{2}
\left[\kappa_\rho X^{-1\mu\beta}(g_{\mu\nu}-f_{\mu\nu})+\kappa_\nu
X^{-1\mu\beta}(g_{\mu\rho}-f_{\mu\rho})-\kappa_\mu X^{-1\mu\beta}(
g_{\nu\rho}-f_{\nu\rho})\right],
\end{eqnarray}where $\christoffelX{\beta}{\nu}{\rho}$ is the Christoffel symbols of the metric
$X$. If we set $X^\mu_{\,\,\,\nu}=\theta^{\mu}_{\,\,\,b}
\sigma^{\,\,\,b}_{\nu}$ \footnote{This and some other useful
relations that may be used in the following calculations can be
obtained sequentially from $X_{\mu\nu}=\theta^{\,\,\,a}_{\mu}
\sigma_{\nu a}$. One can show that
$(X^{-1})^{\mu\beta}g_{\mu\nu}=X_{\nu}^{\hspace{2mm}\beta}$, where
$X^{\hspace{2mm}\beta}_\nu = \sigma^\beta_{\,\,\,b}
\theta^{\,\,\,b}_{\nu}$. Also,
$(X^{-1})^{\mu\beta}f_{\mu\nu}=X^{\beta}_{\hspace{2mm}\nu}$ where
$X^\beta_{\hspace{2mm}\nu}=\theta^{\beta}_{\,\,\,b}
\sigma^{\,\,\,b}_{\nu}$. In addition:
$(g^{-1})^{\mu\beta}X_{\mu\nu}=X^{\beta}_{\hspace{2mm}\nu}$,
$(f^{-1})^{\mu\beta}X_{\mu\nu}=X^{\hspace{2mm}\beta}_{\nu}$,
$(f^{-1})^{\mu\nu}g_{\mu\beta}=X^{\hspace{2mm}\mu}_{\beta}X^{\hspace{2mm}\nu}_{\mu}$
and
$(g^{-1})^{\mu\nu}f_{\mu\beta}=X_{\hspace{2mm}\mu}^{\nu}X_{\hspace{2mm}\beta}^{\mu}$.},
the above relation can be simplified as

\begin{equation}\label{AQ}
\Gamma^\beta_{\nu\rho}=\christoffelX{\beta}{\nu}{\rho}
-\frac{A}{2}\left[\kappa_{\rho}(X_{\nu}^{\hspace{2mm
\beta}}-X^{\beta}_{\hspace{2mm \nu}})
+\kappa_{\nu}(X_{\rho}^{\hspace{2mm \beta}}-X^{\beta}_{\hspace{2mm
\rho}})-\kappa_\mu (X^{-1})^{\mu\beta}(g_{\rho \nu}-f_{\rho
\nu})\right].
\end{equation}
By repeating of the same procedure for the nonmetricity conditions
of the metrics $g$ and $f$, we will obtain their corresponding
connections as

\begin{equation}\label{AR}
\Gamma^\beta_{\nu\rho}=\christoffelg{\beta}{\nu}{\rho}
-M\left[\kappa_{\rho}
\delta^\beta_\nu+\kappa_{\nu}\delta^\beta_\rho-\kappa_\mu
g^{\mu\beta}g_{\rho \nu}\right]+A\left[\kappa_{\rho}
X^\beta_{\hspace{2mm}\nu}+\kappa_{\nu}X^\beta_{\hspace{2mm}\rho}-\kappa_\mu
g^{\mu\beta}X_{\rho \nu}\right],
\end{equation}and

\begin{equation}\label{AS}
\Gamma^\beta_{\nu\rho}=\christoffelf{\beta}{\nu}{\rho}
+M\left[\kappa_{\rho}
\delta^\beta_\nu+\kappa_{\nu}\delta^\beta_\rho-\kappa_\mu
f^{\mu\beta}f_{\rho \nu}\right]-A\left[\kappa_{\rho}
X^{\hspace{2mm}\beta}_{\nu}+\kappa_{\nu}X^{\hspace{2mm}\beta}_{\rho}-\kappa_\mu
f^{\mu\beta}X_{\rho \nu}\right].
\end{equation}Our set-up for writing the geometrical properties of
the conformal group is now complete. In the next section we will
investigate how a gravitation theory may be extracted from this
set-up.

\section{Conformal group as an arena for extension of massive gravity}
An active topic of research in gravity is modifying Einstein-Hilbert
to a theoretically consistent model of gravity. By theoretically
consistent we mean absence of any instabilities including ghosts.
The construction of  modified gravity models are interesting by
their own but they can also be checked with the cosmological
observations. We should recall that a scalar and a vector field
cannot explain the gravity force because of the absence of
interaction with light and anisotropic behavior respectively. Hence
the first chance is a tensor field. It is well-known, according to
Lovelock theorem, that a theoretically consistent theory of a
massless spin-2 particle (graviton) is the unique Einstein-Hilbert
action. This means any modified gravity should have additional
degrees of freedom e.g. scalar-tensor or scalar-vector-tensor
theories\footnote{Note that any other approaches like
extra-dimensions, non-locality can be rewritten equivalently as
Einstein-Hilbert with additional degrees of freedom.}. A very
beautiful extension of gravity is massive gravity which has five
degrees of freedom instead of two for massless graviton
\cite{massive}. On the other hand, as we mentioned Poincar\'{e}
group corresponds to Einstein-Hilbert model and it is natural to
think any more complicated gauge group with additional symmetry
generators should be corresponded to a gravity model with more
degrees of freedom.

In this section, we discuss the question of whether a theory of
gravity could correspond to the above geometric structure. To do
this, we need to assign an action to the model described above. In a
gauge theory point of view, such an action is usually in the form of
the Yang-Mills type actions. In \cite{Lara}, for example, a
Yang-Mills type theory of gravity is written in the framework of the
biconformal gauging of conformal group. Here, we do not do so, but
instead are looking for a gravity theory beyond the Einstein-Hilbert
action in the framework of the geometric structure of the conformal
group. However, in the presence of three metrics which are not
independent according to (\ref{AQ})-(\ref{AS}), the process of
writing the action may be ambiguous. As the first candidate, we may
think of an action which is built by summing up Einstein-Hilbert
actions for every three metric, that is
\begin{equation}
\int \left[M_g^2\sqrt{-g} R(g)+M_f^2 \sqrt{-f} R(f)+M_X^2\sqrt{-X}
R(X)\right]d^4x.
\end{equation}
The above action is not unique in the sense that one can suggest
another types of models. To give a clue that what we can do with
$SO(4,2)$ geometric structure, we propose a simpler model which
still is non-trivial as
\begin{equation}\label{Lag-R1}
{\cal S}=M_g^2\int\sqrt{-g} {\cal R}(\Gamma)d^4x,
\end{equation}where $\Gamma$ is defined in (\ref{AR}). Note that with the above assumption we break the symmetry between metrics $g$ and $f$. By writing
equation (\ref{AR}) in the form of
\begin{equation}
\Gamma^{\beta}_{\nu\rho}=
\christoffelg{\beta}{\nu}{\rho}+C^{\beta}_{\nu\rho},
\end{equation}
we can deduce the Riemann tensor as
\begin{eqnarray}\label{AT}
{\cal
R}^{\beta}_{\nu\rho\alpha}&=&R^{\beta}_{\nu\rho\alpha}+\partial_\rho
C^\beta_{\nu\alpha}-\partial_\alpha
C^\beta_{\nu\rho}+C^{\beta}_{\mu\rho}C^{\mu}_{\nu\alpha}-C^{\beta}_{\mu\alpha}C^{\mu}_{\nu\rho}\nonumber\\
&+&\christoffelg{\beta}{\mu}{\rho}C^\mu_{\nu\alpha}+\christoffelg{\mu}{\nu}{\alpha}C^\beta_{\mu\rho}-\christoffelg{\beta}{\mu}{\alpha}C^\mu_{\nu\rho}
-\christoffelg{\mu}{\nu}{\rho}C^\beta_{\mu\alpha},
\end{eqnarray}
from which the Ricci scalar reads
\begin{eqnarray}\label{Lag-R}
{\cal R}&=&R+g^{\alpha\nu}\left[\nabla_\beta
C^\beta_{\nu\alpha}-\nabla_\alpha C^\beta_{\nu\beta}\right]
+g^{\alpha\nu}\left(C^{\beta}_{\mu\beta}C^{\mu}_{\nu\alpha}-C^{\beta}_{\mu\alpha}C^{\mu}_{\nu\beta}\right),\nonumber
\end{eqnarray}where $R^{\beta}_{\nu\rho\alpha}$ and $R$ are the Riemann tensor and Ricci scalar correspond
to the $g$'s Christoffel symbol. Also, we used the metric
$g_{\mu\nu}$ to take traces and the covariant derivatives,
$\nabla$s, which are compatible with metric $g_{\mu\nu}$. If now, we
return to the definition of $C^{\beta}_{\nu\rho}$, the last term in
the Ricci scalar takes the form
\begin{eqnarray}\label{AU}
g^{\alpha\nu}\left(C^{\beta}_{\mu\beta}C^{\mu}_{\nu\alpha}-C^{\beta}_{\mu\alpha}C^{\mu}_{\nu\beta}\right)
=&-&6M^2\kappa^2 +4A\,M\,\left(\kappa^2
X^\alpha_{\hspace{2mm}\alpha}-
\kappa^\mu\,\kappa^\nu\,X_{\mu\nu}\right)
\\&-&(A^2\kappa^2)\big(X^\alpha_{\hspace{2mm}\alpha}X^\beta_{\hspace{2mm}\beta}-X^\beta_{\hspace{2mm}\alpha}X^\alpha_{\hspace{2mm}\beta}\big)
+2A^2 \big(\kappa^\mu \kappa^\nu
X_{\mu\nu}X^\alpha_{\hspace{2mm}\alpha}- \kappa^\mu \kappa^\beta
X_{\mu\nu} X^\nu_{\hspace{2mm}\beta}\big)\nonumber\\\nonumber
=&-&6M^2\kappa^2+2\,A\,M\,\epsilon^{\rho\sigma\mu\alpha}\epsilon_{\rho\sigma\nu\beta}\kappa_\mu\kappa^\nu
X_{\hspace{2mm}\alpha}^\beta-A^2\,\epsilon^{\gamma\mu\rho\alpha}\epsilon_{\gamma\nu\sigma\beta}\kappa_\mu\kappa^\nu
X_{\hspace{2mm}\rho}^\sigma X_{\hspace{2mm}\alpha}^\beta.
\end{eqnarray}
Hence, the action (\ref{Lag-R1}) up to a total derivative can be
written as
\begin{eqnarray}\label{AV}
{\cal S}=M_g^2\,\int d^4x\, \sqrt{-g}
\bigg[R-6M^2\kappa^2+2\,A\,M\,\epsilon^{\rho\sigma\mu\alpha}\epsilon_{\rho\sigma\nu\beta}\kappa_\mu\kappa^\nu
X_{\hspace{2mm}\alpha}^\beta-A^2\,\epsilon^{\gamma\mu\rho\alpha}\epsilon_{\gamma\nu\sigma\beta}\kappa_\mu\kappa^\nu
X_{\hspace{2mm}\rho}^\sigma X_{\hspace{2mm}\alpha}^\beta\bigg].
\end{eqnarray}
The above action is a kind of vector-tensor model with massive
gravity building block i.e.
$X_{\hspace{2mm}\nu}^\beta=\sqrt{g^{\beta\mu}f_{\mu\nu}}$ where $g$
is dynamical field and $f$ is auxiliary one\footnote{By definition,
if one has $g^{\beta\mu}f_{\mu\nu}= X_{\hspace{2mm}\mu}^\beta
X_{\hspace{2mm}\nu}^\mu$, then
$\sqrt{g^{\beta\mu}f_{\mu\nu}}=X_{\hspace{2mm}\nu}^\beta$. }.
Interestingly, without any fine-tuning, we get the double epsilon
structure which is a hint for ghost-freeness of our model same as
what happens in Lovelock theories, Galileon \cite{galileon} and
massive gravity \cite{massive}. It is worth to say few words why the
double epsilon structure is a hint for ghost-freeness of our model.
Assume a scalar field, $\pi$, with the following Lagrangian
\begin{eqnarray}\label{galileon}
{\cal
L}\propto\epsilon^{\kappa\mu\rho\alpha}\epsilon_{\eta\nu\sigma\beta}
{\big(\partial_\kappa\partial^\eta\pi\big)}{\big(\partial_\mu\partial^\nu\pi\big)}
{\big(\partial_\rho\partial^\sigma\pi\big)}{\big(\partial_\alpha\partial^\beta\pi\big)}.
\end{eqnarray}
For the first look the above Lagrangian will result in higher order
derivative terms (i.e. more than second order) at the level of
equations of motion. This means for solving the equation of motion
one needs more than two initial conditions (i.e. $\pi(0)$ and
$\dot{\pi}(0)$) which means the model contains more than one degree
of freedom. Usually these unwanted degrees of freedom are ghost and
make the model unstable by having a wrong sign of their kinetic
term. However, the above Lagrangian has a very special form because
of the double epsilon structure. A typical term of equation of
motion can be written as
\begin{eqnarray}\label{galileon-eom}
{\cal E}=\frac{\delta {\cal L}}{\delta
\pi}\ni\partial_\gamma\partial^\zeta\frac{\partial {\cal
L}}{\partial
(\partial_\gamma\partial^\zeta\pi)}\ni&&\partial_\gamma\partial^\zeta\bigg
[\epsilon^{\gamma\mu\rho\alpha}\epsilon_{\zeta\nu\sigma\beta}{\big(\partial_\mu\partial^\nu\pi\big)}
{\big(\partial_\rho\partial^\sigma\pi\big)}{\big(\partial_\alpha\partial^\beta\pi\big)}\bigg]=\\\nonumber
&&\epsilon^{\gamma\mu\rho\alpha}\epsilon_{\zeta\nu\sigma\beta}\,\partial_\gamma\partial^\zeta\bigg
[{\big(\partial_\mu\partial^\nu\pi\big)}{\big(\partial_\rho\partial^\sigma\pi\big)}{\big(\partial_\alpha\partial^\beta\pi\big)}\bigg],
\end{eqnarray}
where it is obvious from the last term that if we move a partial
derivative (i.e. one of the $\partial_\gamma$ or $\partial^\zeta$)
into the parentheses then the totally anti-symmetric structure of
one of the epsilons (i.e. one of the
$\epsilon^{\gamma\mu\rho\alpha}$ or
$\epsilon_{\zeta\nu\sigma\beta}$) makes this term zero. This shows
that the double epsilon structure can kill the new unwanted degrees
of freedom. As we mentioned above exactly, this structure makes
Lovelock theories ghost-free even when they are higher order.

We should emphasize that in our scenario we could get massive
gravity terms up to ${\cal{O}}(X^2)$ and we need a kind of
generalization if we want to have all orders of massive gravity
terms e.g. working with another tetrad postulate or assuming higher
order curvature in (\ref{Lag-R1}). The above Lagrangian has been
deduced (more or less) in a bi-connection framework \cite{nima-bi}
where two connections satisfied Weyl nonmetricity with opposite
signs.

Another issue in the above Lagrangian is the absence of kinetic term
for the vector field $\kappa_\mu$. This means our vector field is
not propagating. However, one can add standard kinetic term for a
vector i.e. $F_{\mu\nu}F^{\mu\nu}$  by hand where
$F_{\mu\nu}=\nabla_\mu\,\kappa_\nu-\nabla_\nu\,\kappa_\mu$. A
natural way to add the kinetic term for $\kappa_\mu$ is starting
with higher order gravity models instead of (\ref{Lag-R1}) as it has
been discussed in \cite{Haghani:2014zra}. But we should notice that
since our Lagrangian (\ref{AV}) is not $U(1)$ gauge invariant then
it may suffer from a ghost. Another more natural scenario can be
deduced by introducing a scalar field via $\partial_\mu \phi\equiv
\sqrt{6}\,M\,\kappa_\mu$. In this case the Lagrangian becomes
\begin{eqnarray}\label{AW}
{\cal
L}=R-\partial^\mu\phi\,\partial_\mu\phi+b\,\epsilon^{\rho\sigma\mu\alpha}\,\,\epsilon_{\rho\sigma\nu\beta}\,\,\partial_\mu\phi\,\partial^\nu\phi\,
X_{\hspace{2mm}\alpha}^\beta-\frac{3}{2}\,b^2\,\epsilon^{\gamma\mu\rho\alpha}\,\,\epsilon_{\gamma\nu\sigma\beta}\,\,\partial_\mu\phi\,\partial^\nu\phi\,
X_{\hspace{2mm}\rho}^\sigma \,X_{\hspace{2mm}\alpha}^\beta,
\end{eqnarray}
where $b\equiv A/3M$ is the only free parameter in this model. The
above Lagrangian belongs to quintessence models where the scalar
field is derivatively coupled to the metric via massive gravity
building blocks. As we already mentioned, the additional terms can
be written in double epsilon form. This means the above Lagrangian
should be safe from any ghost in its dynamics. To show this we can
write the most dangerous part of our model in decoupling limit by
replacing $X_{\hspace{2mm}\alpha}^\beta$ by
$\Pi_{\hspace{2mm}\alpha}^\beta=\partial^\beta\partial_\alpha\pi$
where $\pi$ is the scalar degree of freedom of the metric field.
Obviously the double epsilon structure does not allow the equations
of motion have any term with more than two derivatives. This means
our model is ghost free at least in decoupling limit. The above
Lagrangian belongs to a family of massive gravity models with an
additional scalar degree of freedom which has an extensive
literature \cite{quasi}. The more analysis of this model including
its cosmology is beyond the scope of this paper and will be remained
for future works.

\section{Summary}
In this paper we have studied the relation between the Lie groups
and their geometric interpretation. Our interest in this subject is
rooted in the relation between GR and gauge theories, which by a
gauge theory we mean a theory that its action is invariant under a
continuous group which form its symmetry. Our strategy is based on
the building blocks of the Lie groups which show themselves in the
Maurer-Cartan structure equations. These are indeed the objects such
as metric, connection, torsion, curvature, which a geometry may be
equipped with them. The role of the Maurer-Cartan equations is then
to connect them to the generators and gauge fields of the
corresponding Lie group.

The main steps we have taken to address this issue are as follows.
The usual GR as a gauge theory is based on the Poincar\'{e} group.
This group is a semidirect product of two Lie groups: the group of
Lorentz transformations and the group of translations. The
significance of this group becomes apparent when we want to bring
spinors into GR. In order to do so, the usual GR whose connection is
given by the Christoffel symbol compatible with the metric, should
be formulated by the gauge transformation invariance such as
Poincar\'{e} gauge group. For the Poincar\'{e} group we have shown
that while a metric can be constructed by means of the group's gauge
fields the Maurer-Cartan equations equip the resulting metric
geometry with torsion and (Riemann) curvature. Assuming, a tetrad
postulate, we have seen that the metric in this case satisfies the
metricity condition, i.e. its covariant derivative vanishes. This
means that the geometry beyond the Poincar\'{e} group is of the kind
of the Riemann-Cartan geometry.

For the Weyl group, the story is almost the same except that here
the metricity condition is replaced with a nonmetricity relation
which is the characteristic of the Weyl geometry. This group is a
larger group which includes the Poincar\'{e} group as a subgroup.
Here, in addition to the transformations of the Poincar\'{e} group,
a new kind of curvature corresponding to the dilatation, which can
resize the coordinates, has appeared. This means that in the
corresponding gravity theory we have a setting in which
nonmetricity, torsion and curvature are all in play. Then, we have
accomplished our main goal of constructing a geometrical structure
by using of the conformal group which in addition to all
transformations of the Weyl group has an extra generator for the
so-called co-translation. The co-solder forms $\sigma$, the gauge
fields of this extra transformation together with the solder
(tetrad) forms, $\theta$, of the translations opened a new window
for us to build a geometry with three metric. We saw that the
Maurer-Cartan equations of the conformal group, which now included
an additional curvature associated to the co-translations, do not
change under replacements $\sigma \leftrightarrow \theta$ and
$\kappa \rightarrow -\kappa$, where $\kappa$ is the gauge field of
the dilatation. This led us to find the most general nonmetricity
conditions for the metrics from which we deduced the connection.

Finally, we have addressed the issue of whether a gravity theory can
correspond to the geometric structure of the conformal group. Due to
the more complicated structure of the conformal group compared to
the Poincar\'{e}, its gravity theory is expected to have more
degrees of freedom in comparison with the ordinary GR. In this
regard, with the use of the resulting geometry we provided a gravity
theory starting from an Einstein-Hilbert type action whose
Lagrangian is the Ricci scalar of the metric constructed from the
translation's gauge fields. We have shown that the proposed
Lagrangian takes the form of the vector-tensor model with massive
gravity building block. In addition, we rewrote the Lagrangian in a
form which in the massive gravity terminology is called the double
epsilon form. This helped us to show that the extra terms appeared
in the Lagrangian do not include the ghosts, the existence of which
must always be taken into consideration when dealing with such
theories. To show this, we wrote the Lagrangian of a scalar field
with double epsilon structure. In the first opinion, it seems that
such a Lagrangian will result in unwanted degrees of freedom
(ghosts) with wrong sing in the kinetic term which make the
resulting model unstable. However, we explicitly showed because of
the special form of the double epsilon structure these additional
unwanted degrees of freedom will be removed due to the totally
anti-symmetric property of one of the epsilons and thus, the
proposed Lagrangian is safe from any ghost in its dynamics.

\vskip 0.2 in \textit{Acknowledgments:} We would like to thank Tomi
Koivisto for exchanging notes and ideas during the early stage of
this project. N.K. is grateful of School of Physics at IPM where he
is a part-time researcher.

\end{document}